# EQUATION OF STATE OF CHEMICAL SYSTEM: FROM TRUE EQUILIBRIUM TO TRUE CHAOS.

B. Zilbergleyt,

System Dynamics Research Foundation, Chicago

E-mail: livent@ameritech.net

The article presents results of preliminary study of solutions to recently offered basic thermodynamic equation for equilibrium in chemical systems with focus on chaotic behavior. Classical part of that equation was investigated earlier in a series of papers. In this work a similarity between one-dimensional logistic map and non-classical (chaotic) term of the equation was discussed to introduce the problem. Results of this work allow us to evaluate the region where open equilibrium belongs to the basin of regular attractor and leads to trivial solutions with zero deviation from true thermodynamic equilibrium, and then to find first bifurcation threshold as a limit of open equilibrium and a limit of the classical region as well. Features of the basic equation are discussed with regard to relative values of the chaotic and thermodynamic temperatures. Obtained results prompt us to consider the basic equation of new theory to be the general equation of state of chemical systems.

Recently offered basic thermodynamic equation for open chemical equilibrium [1]

$$\ln [\Pi`(\eta_{kj}, 1)/ \Pi(\eta_{kj}, \Delta^*_j)] - \tau_j \Delta^*_j \delta^*_j = 0 \qquad (1)$$

contains ratio of equilibrium constant $K= \ln \Pi`(\eta_{kj},1)$ logarithm to the logarithm of j-reaction mole fraction product, and parabolic, or chaotic term - a product of reduced chaotic temperature $\tau_j = T_{ch}/T_t$, a ratio between chaotic and thermodynamic temperatures, reaction extent $\Delta^*_j$ and reaction shift from true equilibrium $\delta^*_j = 1- \Delta^*_j$. In the numerator of logarithmic expression, reaction extent takes its value for true thermodynamic equilibrium, $\Delta_j =1$. As one can see on Fig.1, all solutions for open equilibrium occur at the crossing of the logarithmic term as function of $\delta_j$ and chaotic term $\tau_j\delta_j(1-\delta_j)$.

The first term in (1) is totally classical - pure $\Delta G_j$ reduced by $RT_t$. The name "chaotic" was assigned to the second term in (1) due to its identity with the right side of so-called *logistic map*

$$x_{n+1} = \lambda x_n(1- x_n), \qquad (2)$$

a good example of equation leading to typical bifurcations and chaos [2].

Values of $\delta^*_j$ in (1) and $x_n$ in (2) are supposed to stay within the range (0,1). Due to intensive study of the chaotic processes for more than two decades, properties of (2) are well known while equation (1), to the best of our knowledge, has never been investigated in this exact form. At the same time, very interesting features of chemical systems may be discovered through its study.

Value of $\lambda$ in the logistic equation, as well as value of $\tau$ in the equation (1) define the fate of iterations, which are important not only for calculating algorithm but also for understanding and control processes in complex chemical systems. We will give a brief summary of key properties of the logistic equation following [2] the refresh the reader's memory. Condition $0<\lambda<4$ and initial choice $x_0(0,1)$ keeps all $x$ within the same range in the run of iterations. If $\lambda<1$, the only steady solution is $x=0$. At $\lambda \approx 3$ 1 first bifurcation occurs and solutions to equation (2) split with period 2. At $\lambda \sim 3.5$ next bifurcation takes place turning the period to 4, at $\lambda \sim 3.54$ period doubles again and becomes equal to 8, and so forth. Further increase of $\lambda$ beyond $\sim 3.5699…$ leads to non-repeating sequence of numbers referred to as *chaotic*.

In case of more complicated equation (1) (OpEq equation), solutions will depend upon relative contributions of both terms – classical logarithmic and non-classical chaotic. It is quite obvious that if on any reason second term equals to zero, chemical system is isolated, and follows totally classical pattern with $\Delta G_j = 0$. This case corresponds to true thermodynamic equilibrium with $\delta_j =0$. The picture will be totally changed if the chaotic term plays essential role in the system behavior, leading it to bifurcations and chaos.

This paper presents results of preliminary investigation of solutions to the OpEq equation using as example some elemental reactions under the influence of these two powers.

Classical paradigm of chemical thermodynamics admits only one state of equilibrium (Zel'dovich's theorem, [3]). This statement is valid only for simple *isolated* chemical system. There are two possibilities in case of open equilibrium before bifurcations and then following chaos occur. The OpEq equation has 2 solutions as minimum – trivial at $\delta=0$, where both curves have one joint point (as with $\tau=1$ on Fig.1), and also non-trivial with $\delta>0$ which represents namely open equilibrium where the terms of equation (1) have at least one crossing point (upper parabola). Both logarithmic and chaotic functions are continuous, differentiable and monotonous, and non-trivial solution of open equilibrium exists if derivative of the logarithmic curve in the initial point $\delta=0$ is less than derivative of the chaotic curve

$$d\ln(\Pi`/\Pi^*)/d\delta|_{\delta=0} < d(\tau\delta\Delta)/d\delta \qquad (3)$$

Minimal value of $\tau$, providing the existence of non-trivial solution, can be easily received from (1). The right side of (3) equals merely to $\tau$, while left side, taking into account that $\Pi`=K$, gives a product $(1/K)\,d\Pi^*/d\delta$, and

$$\tau|_{\delta=0} > (1/K)\,d\Pi^*/d\delta. \qquad (4)$$

Minimal value of $\tau$ is totally defined by K, or the thermodynamic equivalent of chemical transformation $\eta$, and reaction equation (which defines expression for $\Pi^*$). Though numbers used to plot the graph on Fig.1 are taken for example, the result is quite simple - if condition (4) is reversely satisfied, the logarithmic term essentially prevails and open system still has only one attractor - true thermodynamic equilibrium. All states below curve (1) in Fig.1 satisfy condition (4) and solution is $\delta=0$. If a non-trivial solution exists to (1), possibility to find bifurcations due to the chaotic term becomes real.

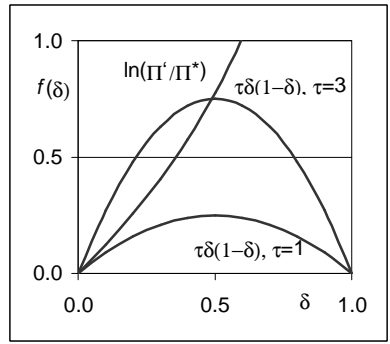

Fig.1. Logarithmic and chaotic terms of the basic OpEq equation as functions of the shift between open and true thermodynamic equilibrium $\delta^*_j$.

Typical iterative graphs are shown on Fig.2. Period n equals to the number of possible solutions.

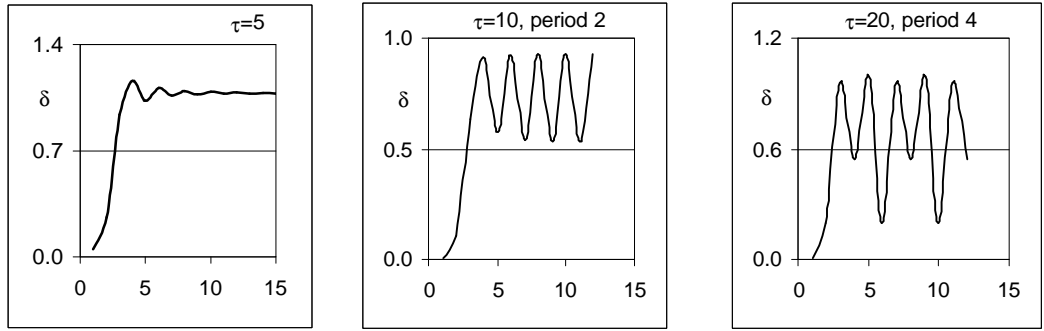

Fig.2. Bifurcation of solutions to the basic equation. Reaction A+B=C, $\eta=0.1$.
Abscissa – number of iterations.

Fig.3 presents state diagram of chemical system with reaction A+B=C. One can distinctively see 3 areas on the diagram – true equilibrium where curves are laying immediately on abscissa and all the way long have $\delta=0$, open equilibrium from the points of $\delta>0$ to the split points, and bifurcations after the split points. At

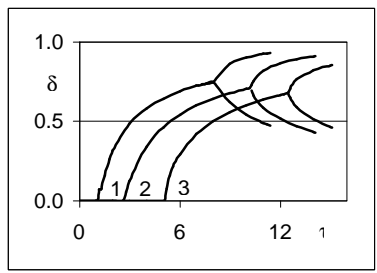

Fig.3. State diagram of chemical system, reaction A+B=C,

corresponding η values: 1 – 0.1, 2 – 0.5, 3 – 0.7.

the tail-ends of the split curves period doubles. Very interesting and important is the fact that open system still stays in open equilibrium up to essential values of reduced chaotic temperatures, and further increase of $\tau$ must occur to move open system up to the split point.

Table 1.

Relation between parameter η and $\tau$ on two upper equilibrium limits: of TTDE – true equilibrium, and of OpEq – open equilibrium, initial reactant amounts equal to unity.

| η | K | (ΔG)/T | Reaction A+B=C | | Reaction A+B=2C | |
|---|---|---|---|---|---|---|
| | | | $\tau$, TTDE | $\tau$, OpEq | $\tau$, TTDE | $\tau$, OpEq |
| 0.1 | 0.24 | 12.06 | 1.2 | 7.4 | 1.3 | 7.6 |
| 0.3 | 1.04 | -0.33 | 1.6 | 8.6 | 1.9 | 8.8 |
| 0.5 | 3.00 | -9.13 | 2.7 | 10.3 | 3.1 | 10.2 |
| 0.7 | 10.11 | -19.24 | 5.2 | 12.5 | 5.7 | 13.5 |
| 0.9 | 99.00 | -38.21 | 13.1 | 18.5 | 15.5 | 20.4 |

Fig. 4 shows explicitly the influence of parameter η which serves as a symbol of the reaction "classical strength". Numerical data is placed in Table 1.

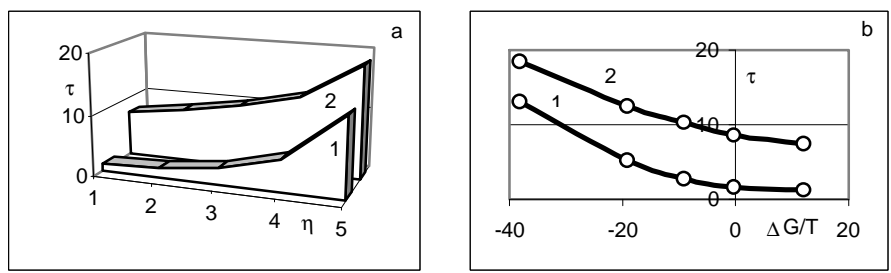

Fig.4. Thresholds of trivial solution (upper limit to true equilibrium, 1) and first bifurcation (upper limit to open equilibrium or equilibrium at all, 2): a) vs. TD-equivalent (points 1 to 5 along x-axis correspond to η=0.1, 0.3, 05, 0.7, 0.9), and b) vs. ΔG/T. Reaction A+B=C, increment of $\tau$ equals to 0.1.

We have placed in Table 1 and on Fig.4b also more habitual values of equilibrium constant K and reduced changes of Gibbs' free energy $\Delta G/T_t$ along with the values of η. Chaotic temperature in all cases as well as change of free Gibbs' energy is reduced by thermodynamic temperature, and both are counted in values per Kelvin's degree.

Due to similarity between the shift-external force graph and the Hooke's law with its yield point [1], OpEq has two values of $\tau$ – below the yield point and above it. To investigate the case with two values of $\tau$ the calculation program was designed to take this feature into account. The results show that if initial approximation $\delta_y$ was within reasonable limits (0.3 – 0.7; the values found earlier for real systems), the second $\tau$ played major role in the system behavior, the two-$\tau$ calculation results didn't bring any difference comparing with one-$\tau$ approach.

We would like to focus reader's attention once again on Fig.4. Area below curve 1 in Fig.4 corresponds to true thermodynamic equilibrium ($\delta$=0) though in this area $\tau$>0. Area between curves 1 and 2 corresponds to open equilibrium (0<$\delta$<1) where non-trivial solutions to the equation (1) reside. The area above curve 2 corresponds to bifurcations and perhaps to no equilibrium at all. Therefore, zone of open equilibrium with non-zero shifts separates true equilibrium and chaos, or classical and non-classical areas, serving as water shed between basins of regular and strange attractors. It is noteworthy and quite natural that the more negative is standard change of free Gibbs' energy of reaction the higher are the limit values of reduced chaotic temperatures. All the areas described above can be explicitly seen on Fig.3.

The picture on Fig.3, which also may be considered an evolution diagram of chemical system, is not new and doesn't need any references. The new is that it represents a diagram of states of chemical system covering in general all conceivable situations, from true equilibrium to true chaos, and that its particular points depend upon reaction change of free Gibbs' energy and upon external thermodynamic force as well. Also new is a conclusion that states of virtually every chemical system, no matter how simple is it's "signature" in terms of chemical reaction running within the system, may be described by similar diagram.

A subtitle to this paper could be "*How simple is simple reaction A+B=C?*". As it follows from here presented preliminary results, this reaction ceases to be simple as soon as it runs in open chemical system and may fall above the area with zero deviations from classical equilibrium.

Classical thermodynamics of chemical equilibrium deals with chemical reactions, formally uniting them into a set if their number exceeds one. Non-classical thermodynamics also operates reactions, even if they are designed artificially (e.g., brusselator, oregonator, etc.). Discerning feature of the theory we discuss in this article is that it consistently handles chemical systems, not reactions.

Basic equation of new theory links equilibrium (corresponding to isolated systems) and non-equilibrium thermodynamics (making sense in open systems), and may be rewritten more generally as

$$\Delta G_j = \Delta G^0_j + RT_t f_t (\Delta^*_j) + RT_{ch} f_{ch} (\Delta^*_j). \qquad (25)$$

The method treats true, isolated thermodynamic equilibrium of a system as a reference state for its open equilibrium when system becomes a part of another system. This reference state is memorized in $\eta_{kj}$. Such an approach is well fitting the interpretation of equilibrium at zero control parameters as origin of the chaosity scale (the S-theorem, [5]) with one difference – we have a whole area, surrounding zero point of the reference scale instead of a single initial point. That means that in the case of chemical system a whole *zero response* area (below curve 1 on Fig.4b) is essential instead of a point with zero control parameters. Based on a very simple and quite natural assumption, *equation of state of open chemical system naturally and smoothly drags non-linearity into thermodynamics of open systems, unifying states of true equilibrium, open equilibrium and areas of bifurcations and chaos and bridging a gap between classical and non-classical thermodynamics. Offered in this paper theory represents new, unified thermodynamics of chemical equilibrium fitting open systems as well as isolated chemical systems in a particular case.*

One can see that behavior of open system is essentially affected by parameter $\tau$. Depending upon the value of standard change of Gibbs' energy (or $\eta$), system still will not deviate from true, classical thermodynamic equilibrium if external impact doesn't move the ratio $T_{ch}/T_t$ beyond a certain value. The bifurcation threshold features the same dependency. One may strongly declare *that evolution of open chemical system from thermodynamic, "dead" order through bifurcations to "vivid" chaos, i.e. its transition from kingdom of thermal energy to the point where it gives up to external power, is driven by ratio between chaotic and thermodynamic temperatures.*

The chaotic temperature is a parameter of new theory. At the same time we should confess that so far our actual understanding of its physical meaning is not clear. The value of reduced chaotic temperature $\tau$ can be easily found in open equilibrium for any equilibrium value of reaction extent $\Delta^*$ directly from the basic equation (1) and may be immediately used in thermodynamic simulation. Also, the state diagrams of open systems may be used to find correspondence between values of $\delta$ and $\tau$. This makes new theory and following from it Method of Chemical Dynamics (MCD, [1]) available for practical needs. The advantage of the method of chemical dynamics is not restricted by opportunity to avoid using coefficients of thermodynamic activity. The method, for instance, also leads to new in principle opportunity to simulate internal equilibrium of a system with subsystems at different thermodynamic temperatures (like in plasma).

To conclude we would like to mention that open chemical system is by definition coupled with another open system, more exactly, with its compliment to a bigger system. Changing the control parameters may cause an adjustment of the whole system to new equilibrium state through the bifurcations area, and one will observe it as a system of coupled oscillators [2].

The new theory gives alternative, dynamic approach to chemical systems as opposite to conventional probabilistic approach [4], and offers unified concept of chemical system where true and open equilibrium, bifurcations and chaos logically tied up together by unknown before equation of state of chemical system.

Some other related problems and calculation results can be found in [6].

**REFERENCES.**